\documentclass[preprint]{vgtc}               
\ifpdf
  \pdfoutput=1\relax                   
  \usepackage{graphicx}                
  \DeclareGraphicsExtensions{.pdf,.png,.jpg,.jpeg} 
\else
  \usepackage{graphicx}                
  \DeclareGraphicsExtensions{.eps}     
\fi%

\graphicspath{{figures/}{pictures/}{images/}{./}} 

\usepackage{microtype}                 
\PassOptionsToPackage{warn}{textcomp}  
\usepackage{textcomp}                  
\usepackage{mathptmx}                  
\usepackage{times}                     
\usepackage{cite}                      
\usepackage{tabu}                      
\usepackage{booktabs}                  
\usepackage{gensymb}
\usepackage{amsmath}

\onlineid{2005}

\vgtccategory{Full Paper}

\vgtcinsertpkg

\preprinttext{This is the author's version, to appear in the IEEE VR 2025 conference.}

\title{Spatial Bar: Exploring Window Switching Techniques\\for Large Virtual Displays}

\author{Leonardo Pavanatto$^{1}$\thanks{lpavanat@vt.edu} , Jens Grubert $^{2}$, Doug A. Bowman$^{1}$}

\affiliation{\scriptsize $^1$ Center for Human-Computer Interaction, Virginia Tech, USA \\ \scriptsize $^2$  Coburg University of Applied Sciences and Arts, Germany  %
}

 \teaser{
   \centering
   \includegraphics[width=\linewidth]{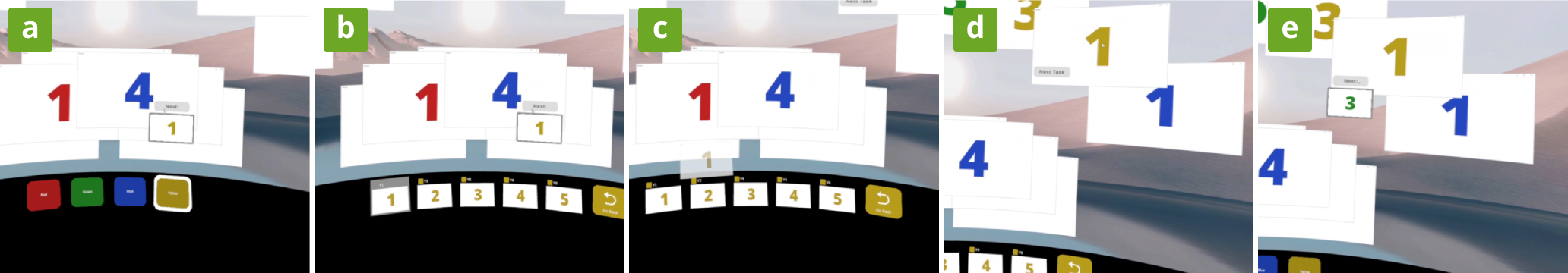}
   \caption{Switching windows with the Spatial Bar: (a) user highlights desired category and clicks on the input device to confirm selection; (b) windows on category open, and the user selects the desired window; (c) animation shows where the window is located; (d) cursor teleports to the center of the window; (e) user clicks a button to end task and show next target.}
   \label{fig:teaser}
 }
\abstract{
Virtual displays provided through head-worn displays (HWDs) offer users large screen space for productivity, but managing this space effectively presents challenges. Existing research shows that while increased screen space improves performance, it can also introduce significant window management overhead. This paper explores how to enhance window-switching strategies for virtual displays by leveraging eye tracking provided by HWDs and underutilized spaces around the main display area. We investigate the efficiency and usability of different cursor behaviors and selection modes in a Spatial Bar interface for window-switching tasks in augmented reality environments. Our study involved two primary selection modes: \textsc{Gaze} and \textsc{Cursor}, each tested with two cursor behaviors: \textsc{Teleport} and \textsc{Stay}. We measured objective performance metrics, including task completion time and error rates, and subjective evaluations using the NASA TLX and custom questionnaires. Results show that \textsc{Cursor Stay}, while familiar and comfortable for participants, was laborious and led to longer task completion times, particularly over large distances. \textsc{Gaze Teleport} led to the quickest window switching times, particularly in tasks where the original cursor position or the target window were far from the Spatial Bar.

} 

\CCScatlist{
  \CCScatTwelve{Human-centered computing}{Human computer interaction (HCI)}{Interaction paradigms}{Virtual reality};
  \CCScatTwelve{Human-centered computing}{Human computer interaction (HCI)}{Interaction paradigms}{Empirical studies in interaction design}
}

\begin{document}

\firstsection{Introduction}
\maketitle
Research has indicated that having more screen space can lead to performance improvements when conducting productivity work \cite{czerwinski_toward_2003} and that such space can be provided through virtual displays \cite{pavanatto_multiple_2024, pavanatto_virtual_2024}---flat or curved surfaces that can display the output of a personal computer at any size through a head-worn display (HWD). Other works have explored how such displays can be used in transportation \cite{medeiros_shielding_2022} and remote work settings \cite{pavanatto_working_2024}, often using a laptop, and the effects of working with them for a week \cite{biener_quantifying_2022}. One downside of adopting such displays, however, is that if the user has more space available, it may become more difficult to manage that space---the extra space can introduce overhead in performing window management operations \cite{pavanatto_multiple_2024}.
Pavanatto et al. \cite{pavanatto_multiple_2024} have shown that providing users with a large virtual space does not mean they will use it well. They observed that a single canvas display led to windows being positioned closer together at the cost of more window management. In contrast, a multi-monitor display with boundaries made windows more spread out, with space being wasted with unuseful information. Based on such results, we cannot expect users to efficiently manage such a large space with existing window management interfaces. Large virtual displays need efficient techniques for rapidly switching among a large number of windows.

Regarding window-switching strategies, common mechanisms such as using a taskbar or dock may be limited. On a canvas display, that implies the user has to move their cursor from its current location to the bottom center of the screen to reach the taskbar, find and select their target window, find out where the window was restored to on the large screen, and move the cursor to the new window. This step becomes more complicated in multi-monitor setups or when using multiple virtual desktops, where the user must determine on which monitor the window will appear or which desktop contains that window. While another popular strategy using a keyboard shortcut (e.g., ALT+TAB), can reduce cursor movement by placing windows on a centered pop-up and cycling through key presses, it doesn't scale as well to a large number of windows. It will require repeated key presses until cycling through the right window or cursor movement to select it directly (in modern implementations). An exposé or WIN+TAB approach will further complicate usage on large screens, as they spread windows across the entire monitor space, requiring large cursor movements. Those strategies work very well for traditional monitors, but when we consider large displays with a large number of windows open, it can take time for the user to reach and retrieve their window of interest---as they would be performing a linear search. We believe the flexibility of virtual displays can provide us with opportunities to improve such strategies.

Two characteristics of virtual displays are useful in supporting the design of new window-switching interfaces. We first make use of \textit{Underutilized Spaces}. Those are spaces around a monitor that are not used by the monitor itself---for instance, the space around the monitor, attached to the user's body (e.g., the forearms), on the desk beside the keyboard on the side opposite the mouse, and even behind the monitor (such as near a wall that is far in the back). These spaces cannot be utilized by monitors constrained to a 2D surface due to occlusion, different orientations, being near other physical objects, and similar constraints. But they can hold smaller virtual interfaces supporting users in their daily work, such as widgets, tools, and individual windows \cite{lu_--wild_2023}. We take advantage of underutilized space to place a taskbar-like interface outside the main display. While the monitor underutilizes the space, we note that the user might still have to negotiate the placement of other physical or virtual objects in the space. Second, we make use of the idea of combining traditional input devices and interaction metaphors with advanced sensors provided by the HWD to improve cursor movements \cite{ashdown_combining_2005, benko_pointer_2007}. For instance, Biener et al. \cite{biener_breaking_2020} have used head tracking to support window selection. Others have used eye tracking complemented by a confirmation input, such as gaze + pinch \cite{pfeuffer_gaze_2017, mutasim_pinch_2021}. We adapt this approach to \textit{gaze+click} since we assume that large virtual displays already have access to standard computer input devices such as a mouse or trackpad.

In this work, we analyze window management on a virtual display using a trackpad for interaction. We propose a technique called the \textit{Spatial Bar}, seen in \autoref{fig:teaser}. Thumbnails of all open windows are displayed below the display. Coupled with gaze+click, users can look at the thumbnail of the window they want to restore and click on an input device to confirm the action. This removes the need for the user to move the cursor a large distance to interact with the Spatial Bar interface. In addition, based on the assumption that the user wants to interact with the newly activated window, our technique can teleport the cursor to the center of that window after a selection in the Spatial Bar. This removes the need for the user to remember where the cursor was located previously or to move it to the location of the new window. While we believe this technique could be used for either virtual or physical monitors and with a mouse or trackpad, we focused on the use case of workers in remote settings, who often use a laptop \cite{ng_passenger_2021, pavanatto_working_2024}.

We conducted a user study to understand the effects of two components of the Spatial Bar technique on window-switching performance, workload, and user experience. The \textit{Method of Selection} could be \textsc{Cursor} or \textsc{Gaze}, while the \textit{Teleportation Behavior} could be \textsc{Stay}, where the cursor does not move after a window selection, or \textsc{Teleport}. The \textsc{Cursor-Stay} combination is similar to a standard taskbar, while the \textsc{Gaze-Teleport} combination is our novel Spatial Bar technique. We measured time, accuracy, cognitive load, and perceived ratings. Results showed that \textsc{Gaze} can be faster than \textsc{Cursor} when switching between windows at a large distance from the Spatial Bar, while \textsc{Cursor} can be faster than \textsc{Gaze} when windows are at a close distance.

\section{Related Work}
Previous research on large monitors has shown that when conducting cognitively difficult tasks, a display with larger screen space can provide a significant advantage in performance \cite{czerwinski_toward_2003, cetin_visual_2018}. Enhanced performance was attributed to physical navigation and maintaining an overview context \cite{ball_effects_2005}. The location and visual appearance of content in large displays also become valuable clues to keep users aware of the organization of the space as a type of external memory \cite{andrews_space_2010, ball_move_2007}. These findings indicate the importance of having more screen real estate and accessing it using body motion, also highlighting some limitations of using current window or desktop switching approaches.

Working with large monitors can also introduce issues such as losing track of the mouse cursor, accessing distant information/windows, dealing with bezels, and managing windows and tasks in the extra and distant available space \cite{robertson_large-display_2005, endert_designing_2012}. Deciding where to place a new window, how to quickly move it, and how to organize a space with a large number of windows are some window management issues \cite{robertson_large-display_2005}. Differences in physical configurations can also impact the usage of such systems, where the designer needs to carefully consider placement strategies for the mouse, keyboard, and displays \cite{endert_designing_2012}. These factors may also impact large virtual displays, although some of them can be addressed more easily without the physical constraints.

Conducting productivity work in HWDs can provide more display flexibility, reducing costs \cite{pavanatto_monitors_2021} and addressing many challenges such as lack of space, illumination issues, and privacy concerns \cite{grubert_office_2018, ofek_towards_2020}. VR has reduced the distraction of users working in open office environments, induced flow, and was preferred by users \cite{ruvimova_transport_2020}. However, it can also force additional head rotation that could result in neck pain \cite{grubert_office_2018}, which could be minimized by the use of amplified head rotation (i.e., performing a virtual movement opposite to the physical head movement) \cite{mcgill_expanding_2020}. Organizing displays around a cylinder has been a strategy to display elements closer to the user, with commercial tools such as BigScreen \footnote{https://www.bigscreenvr.com/} and Virtual Desktop \footnote{https://www.vrdesktop.net/}. Wei et al. \cite{wei_reading_2020} further showed that text should only be warped in a single direction at a time and at small curvatures to preserve reading comfort. This indicates that large virtual displays could bring benefits to productivity work, although more research on the window management aspect is needed.

Existing work has explored some ways of positioning and organizing content in virtual displays. Ens et al. \cite{ens_personal_2014} investigated the design and advantages of an immersive multitasking system that supports surrounding the user with windows. They called those 2D planar elements ``Ethereal Planes,'' \cite{ens_ethereal_2014} and categorized them in seven dimensions: perspective (egocentric, exocentric), movability (can it move?), proximity (on body, near, far), input mode (direct, indirect), tangible (is it tangible?), visibility (visible, partially, no visuals), and discretization (discrete, continuous). Combining virtual displays with laptop/tablet touchscreens was a feasible approach to aid mobile workers while using strategies such as head tracking for selecting which window to focus on \cite{biener_breaking_2020}. Combining virtual monitors with tablets used for touch input has also been shown to achieve performance and accuracy comparable to touch-controllers \cite{le_vxslate_2021}. Despite these advances, there is still no solution for rapid window switching on large displays in standard desktop computing settings.

While not used for specifically for window management, we were also inspired by interaction techniques that combine the use of gaze with pinch \cite{pfeuffer_gaze_2017} and with touch input on tablets \cite{pfeuffer_gaze-touch_2014, pfeuffer_gaze_2016}---to provide users with fast confirmation to gaze selection. However, existing solutions do not perform direct comparisons to understand how they compare to traditional approaches. 

Our work addresses these gaps by introducing the Spatial Bar, combining the use of underutilized spaces, gaze+click selection, and cursor teleportation in order to improve window switching efficiency on large virtual displays; and by conducting a controlled user study to validate our approach.

\section{Spatial Bar Technique}
\begin{figure}[tb]
 \centering
 \includegraphics[width=\columnwidth]{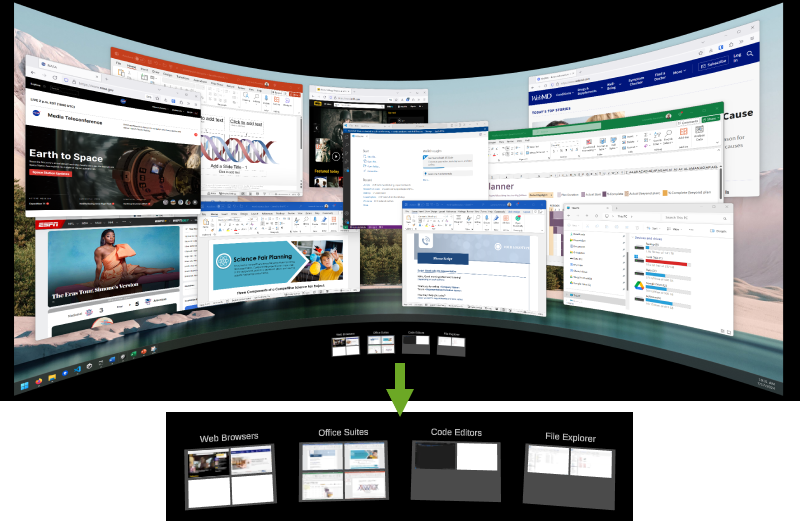}
 \caption{Spatial Bar is shown below the virtual display. Categories can be seen in the interface; in this case, they were Browser, Office, Coding, and Explorer.}
 \label{fig:ch8_1_spatialBar}
 \vspace{-5mm}
\end{figure}

The Spatial Bar is positioned immediately below the computer display, between the screen and the keyboard---allowing users to focus on their current work, rather than cluttering the screen space. It displays thumbnails of all open windows in the system, independently of whether they are visible, minimized, occluded, or out of the user's field of view. The thumbnails are organized in categories, allowing many windows to be represented without taking too much space, and allowing thumbnails to be large enough to be understood at a glance. Windows can be categorized in multiple ways, such as active task, type of application, or simply software name. An image of the Spatial Bar can be seen in \autoref{fig:ch8_1_spatialBar}.

To switch to a target window, the user gazes at the category containing the target and then confirms the selection by clicking on their input device (e.g., trackpad). While the input device is usually used for manipulating the cursor, when the user is looking at a Spatial Bar target, the meaning of the click switches to gaze confirmation. The category then opens to reveal thumbnails of its windows. The user then performs the same action again, this time selecting the window thumbnail. Once confirmed, an animation shows the location of the real window on the display, and the user's cursor automatically teleports to the center of that window. The user then switches back to using the cursor to interact with the window. Those steps are shown in \autoref{fig:teaser}.

We designed the Spatial Bar to provide animations guiding the user to the window's location. Once the user selects a thumbnail, a semi-transparent clone of the thumbnail will move toward the real window while changing size to match the real window. The animations take 0.3s, enough for users to see and follow them, but fast enough that it doesn't slow down their interaction \cite{nielsen_usability_1994}. The main reason for providing such a feature is that virtual displays can have larger sizes, and parts of the display can easily fall out of the field of view. We needed to provide cues to the user so that they understood that their selection was successful and where the window was located. This could also be used for multi-monitor setups; instead of trying to remember which monitor a window is being restored to, the user can follow the animation.

We investigate two critical design choices for Spatial Bar in a user study. For selection, we opted to use a gaze+click technique. This entails that the user will highlight the category or thumbnail they want to select with their eye gaze and then simply click on their input device to confirm the selection---the input device can be a mouse or a trackpad, the same one the user was already using to interact with their computer. Our reasoning behind this choice is that eye gaze should move faster than the cursor when traversing a large space \cite{sibert_evaluation_2000, fono_eyewindows_2005}. By using click for confirmation, we allow users to continue using the same input device, with no need for delays (such as using gaze dwell) or for moving into a different modality (such as pinching or voice commands). However, it could potentially lead to more cognitive load, as users have to mentally switch the meaning of a click, as it is used to confirm both cursor and gaze selections. Then, to help support the user in recovering their context on the display, we used a teleportation approach. As the user selects the thumbnail with their gaze+click, the system automatically moves the cursor to the center of the retrieved window. In that way, users do not need to remember where their cursor was before, which we believe will make it quicker for them to transition between cursor and gaze modalities. A potential downside would be if users do not want or do not expect teleportation to occur, which could lead to extra movement to fix the mistake.

\section{User Study}
We conducted a user study to validate our design choices and obtain user feedback about the Spatial Bar. We were aiming to answer two research questions, ``How does a technique that combines gaze and click compare to a traditional cursor interface in terms of performance (speed and accuracy) during a window-switching task in VR when considering windows at short and large distances?'' and ``How does alternating the meaning of the input click between traditional cursor interaction and gaze selection impact user’s performance and workload in VR, and how it is impacted by teleportation?'' We wanted to understand how well gaze serves users when performing window switches, as we believe they should be faster over the large spaces. As we decided to use the same input device for selection confirmation and normal window interaction, we must also understand if switching the meaning of a click to confirm selections from both cursor and gaze would increase mental workload and lead to confusion.

During this study, we used a virtual monitor rather than a physical one to achieve a larger screen, and to avoid limitations of video pass-through AR. We conducted the study in VR instead of AR to avoid any elements in the real world that could distract the user from the time-intensive task. Finally, we opted to use a trackpad rather than a mouse since our target for virtual displays to work anywhere is laptop hardware, which conventionally uses a trackpad. Moreover, mouse performance can vary considerably given the amount of available physical space and the related need for clutching.

\subsection{Conditions}
Our study included four conditions, which arise from two independent variables with two levels each. We varied the \textit{Method of Selection} in the Spatial Bar (\textsc{Gaze} or \textsc{Cursor}). \textsc{Gaze} refers to using eye tracking to highlight the desired thumbnail by simply looking at it and then clicking on the trackpad to confirm the selection. \textsc{Cursor} refers to using the trackpad to move a cursor and highlight the desired thumbnail, and then click on the trackpad to confirm the selection. We also varied the \textit{Teleportation Behavior} of the cursor (\textsc{Teleport} or \textsc{Stay}). \textsc{Teleport} implies that after selecting a thumbnail in the Spatial Bar, the cursor will automatically teleport to the center of that window in the virtual display---allowing users to just look at the window to see their cursor already there. \textsc{Stay} implies that the cursor will stay or remain in place after a thumbnail selection is performed; it will not teleport---which implies that users will have to move the cursor to the revealed window, and, in the case of Gaze selection, may need to remember the cursor's previous location on the virtual display.

By combining those variables, we have four conditions: \textsc{Gaze-Teleport}, \textsc{Gaze-Stay}, \textsc{Cursor-Teleport}, and \textsc{Cursor-Stay}. \textsc{Gaze-Teleport} is our preferred design for the Spatial Bar, and \textsc{Cursor-Stay} works similarly to traditional operating systems, such as Windows' Taskbar or MacOS' dock. \textsc{Gaze-Stay} and \textsc{Cursor-Teleport} represent middle grounds between them, allowing us to understand the effects of the independent variables on our measures. In all conditions, we used the same organization for the Spatial Bar, with two levels (categories and thumbnails inside those categories), and we used an animation to show where the restored window is located, allowing users to follow it with their eyes and arrive at the target window quickly. Also, across all conditions, we used the same trackpad, positioned in front of the user, and the same virtual display, curved around the user and shown over a black virtual environment. As our task didn't require any interaction with the real world other than moving their fingers and clicking on the trackpad, we preferred to keep them isolated in VR and focused on the task at hand rather than showing a real-world background with an AR display.

\subsection{Implementation and Apparatus}
We simulated a window management system that contained only the operations we required, allowing us to reduce confounding variables and strategy variability between participants. Our simulator was created in the following manner: (1) we created a surface in Unity to be our display background and curved it at 1m radius from the user; (2) we created 20 windows using similar surfaces but of smaller size (1536 by 864px, or 1/5 of the display size), also curved around the user, at the same depth as the background---we used custom shaders to set the rendering order; (3) we created a simulated cursor that would move according to the user input, set to always render on top of all windows; (4) we mapped the cursor onto the curved surface---allowing movements to follow the pixels on the virtual screen; (5) we used a raycast from the cursor to select elements; (6) we allowed the cursor to get out of the screen, over the Spatial Bar, for the \textsc{Cursor} conditions.

We calibrated the cursor speed through a pilot. Three participants completed the subtasks across the conditions, with the option to adjust the cursor speed. Participants were told to balance between moving the cursor fast across large distances, selecting buttons with ease, and reducing overshooting---when the cursor moves further than desired. All participants agreed on a sensitivity of 20 (i.e., the delta cursor value is multiplied by 20). We used the fastest available setting regarding the cursor speed configuration on Windows.

Our implementation used the Unity Engine, version 2021.3.37f1, with a Meta Quest Link for a tethered connection. We used an Apple Magic Trackpad 2 as the input device. We ran the experiment on a computer with an Intel i90-12900K CPU, a Geforce RTX 3070 Ti graphics card, 32GB of RAM memory, and a 1TB SSD. A Meta Quest Pro was used in all conditions. It has a field of view of 108$\degree$ horizontally and 95.57$\degree$ vertically, and a resolution of 1800x1920 per eye. Head and eye tracking were obtained using the Meta All-in-One XR Plugin v64. The virtual monitor was the equivalent of one 8k monitor (7680 x 4320 pixels), with a physical size of about 74 diagonal inches in total, positioned at 1m from the user.

\subsection{Experimental Design}
Our within-subjects study had four independent variables: the \textit{Method of Selection} (\textsc{Gaze} vs \textsc{Cursor}), the \textit{Teleportation Behavior} (\textsc{Teleport} vs \textsc{Stay}), the initial window to spatial bar distance (\textsc{Short} vs \textsc{Large}), and the spatial bar to target window distance (\textsc{Short} vs \textsc{Large}). Presentation order was counterbalanced using a Balanced Latin Square. We recruited 16 participants from the general population who fit the following inclusion criteria: (1) were at least 18 years old, (2) had normal vision (glasses and contact lenses were accepted, but we asked participants not to use them if they could, to reduce issues with eye tracking---we asked them to read small text on the HWD to test; conditions such as color blindness and strabismus were not accepted), (3) were proficient with the English language, and (4) used a computer daily for work. We confirmed the criteria with participants prior to scheduling a session and further screened for color blindness before starting the experiment---inside the HWD, we presented four numbers with four color backgrounds and asked participants to say the number and color pair out loud. Regarding window distances, \textsc{Short} used a distance of 25cm between the center of the window and the center of the Spatial Bar, while \textsc{Large} used 70cm. We included this variable since we expected that the effectiveness of \textsc{Gaze} vs \textsc{Cursor} and \textsc{Teleport} vs \textsc{Stay} might depend on the distance the cursor needed to travel.

Our dependent variables included objective measures of task completion time (in milliseconds) and error rate (number of selection errors), and subjective ratings from questionnaires assessing participants' perception of using each condition (custom questionnaire) and cognitive load (raw NASA TLX). We further performed semi-structured interviews to gather more details about their experience.

\subsection{Hypotheses}
Our hypotheses regarding \textit{Method of Selection} and \textit{Teleportation Behavior} were as follows:

\textbf{H1. \textsc{Gaze} will perform faster and be preferred over \textsc{Cursor} when the previous window location is far from the Spatial Bar.}
We expect that gaze will be faster than cursor because eye movements should be faster than cursor movements \cite{jacob_what_1990}, especially for longer distances within the VR environment, given that users may reach the edge of the trackpad and reposition their hand. We hypothesize that it will also be preferred because users will perceive it as less work and more speed.

\textbf{H2. \textsc{Cursor} will perform faster and be preferred over \textsc{Gaze} when the previous window location is near to the Spatial Bar.}
We believe that the precision and familiarity of cursor interaction may outweigh the speed advantages of gaze interaction for shorter distances. While moving the cursor over large distances can bother users, we expect it will offer more precision and speed in a short space.

\textbf{H3. \textsc{Gaze} will result in a higher selection error rate than \textsc{Cursor}.}
Potential imprecision in eye-tracking technology \cite{feit_toward_2017, fernandes_leveling_2023} can impact task accuracy. We tried to minimize this impact by providing confirmation with the trackpad as a separate device that has little influence on eye position and a short response time, but we still expect it will be easier for the eye tracking to make a mistake and lead to misclicks.

\textbf{H4. \textsc{Gaze-Teleport} will lead to a decrease in task completion time and  workload compared to \textsc{Gaze Stay}.}
Both \textsc{Gaze} and \textsc{Teleport} have the potential to make interaction faster, but we expect that the benefits will be dependent on their pairing. When paired together, we hypothesize they will lead to a decrease in task completion time and workload compared to \textsc{gaze-stay}. Despite the cost of switching from interacting with gaze back to using the cursor, we believe that \textsc{teleport} may decrease this cost as an affordance to mentally guide them back to using the trackpad for moving the cursor.

\textbf{H5. \textsc{Cursor-Stay} will lead to a decrease in task completion time and workload compared to \textsc{Cursor-Teleport}.}
Similarly, we hypothesize that the more traditional approach will work better than performing \textsc{Teleport} on a cursor selection. We believe that if the cursor is already being used for selection, performing teleportation would lead to more confusion and flow interruption, as the mental model gets broken (i.e., sometimes, the cursor behaves continuously, and sometimes, it teleports to another location).

\subsection{Experimental Task}
\begin{figure}[tb]
 \centering
 \includegraphics[width=0.75\columnwidth]{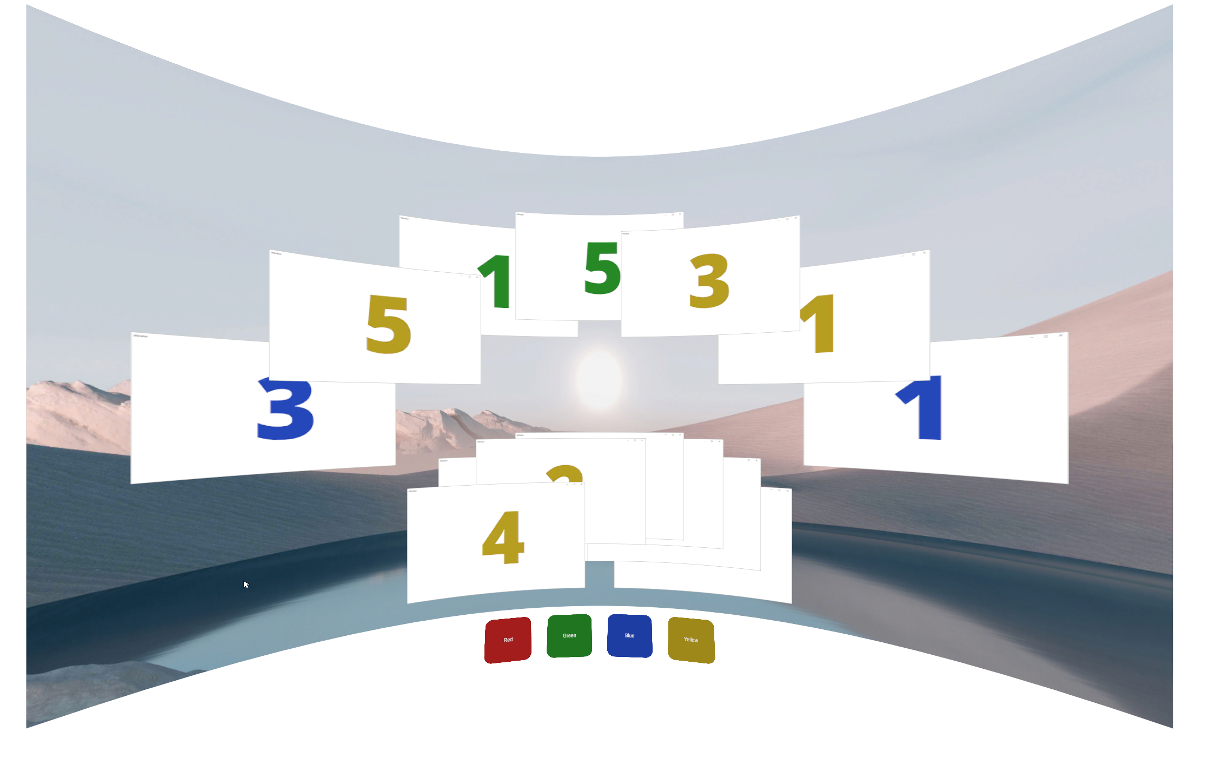}
 \caption{The environment used in the study simulated window management. Windows had colors and numbers, and categories in the spatial bars represented the colors.}
 \label{fig:ch8_2_spatialBar}
  \vspace{-5mm}
\end{figure}

Participants were asked to perform a window management task inside a VR environment. The environment consisted of a virtual display with a Spatial Bar, as can be seen in \autoref{fig:ch8_2_spatialBar}. The display had 20 windows placed in semicircles with two radii (25cm and 70cm). Each window contained only a large number painted in a specific color. We utilized four colors (Red, Green, Blue, and Yellow) and numbers from 1 to 5. We then asked participants to switch to a specific window (e.g., Blue-4) in each subtask. The first subtask in a set (which was during training) always asked participants to select the window with the red number 1. In the Spatial Bar, at the top level, we displayed the categories (which, for this study, were the 4 colors). As a category (such as red) was selected (by either \textsc{Gaze} or \textsc{Cursor}), the items would be replaced by thumbnails of the windows of that color (red numbers from 1 to 5). We further added a "Go Back" button to allow participants to return to the color selection in case of a selection error.

We opted for using simplified windows and color categories to reduce confounds in this study, as it would be evident to users which category they need to go for any given window. In a real-world scenario, users would have to decide how the windows would be categorized and then remember to pick the correct category when restoring the window---from that perspective, our study is similar to a Fitts' Law study \cite{mackenzie_fitts_1992}.

Once a thumbnail was selected, it would bring the corresponding window to the foreground on the virtual display and show a button inside the window. This button had a label "Next task," and participants had to click on it to complete the current subtask and automatically start the next one. This button was selected with the cursor across all conditions. The reasoning was that we were only testing the use of gaze for switching between windows, meaning that users would still interact within each window using their cursor---we did not want to replace the cursor with the user's gaze completely but rather complement it. Furthermore, the position of the button was randomly decided on every trial. The position was determined so that it would be a fixed distance from the center of the window but in a random direction. The position of the button would never intersect with the actual window center, ensuring that participants would always have to move the cursor to press the button, even after teleporting. Once the button was pressed, a picture of the next target window would be shown right below it, making sure participants could quickly see what to do next. The steps are showing in \autoref{fig:teaser}.

In total, each participant performed 60 subtasks (i.e., window switches) per condition (a total of 240 window switches). Of those, the first five were used for training---to explain how to use the conditions and let participants practice. After that, we moved to the main task, where the remaining 55 window switches were performed sequentially---as one subtask ended, the next one started automatically. We discarded the first three window switch operations for all users. This was due to two reasons: the users could make more mistakes at the beginning of the main task, and we wanted the number of subtasks to be a multiple of four. As we considered two distances for the windows, and each window switch operation consists of moving from one window to the spatial bar and then from the spatial bar to the other window, we had four combinations of distances in our subtasks (\textsc{Large-Large}, \textsc{Large-Short}, \textsc{Short-Large}, and \textsc{Short-Short}. By recording results for 52 subtasks, that led to 13 window switches per distance pair.

\subsection{Procedure}
The ethics committee at our university approved the study. The study took place face-to-face at our laboratory in a single session of 60-75 minutes. We recruited participants through mailing lists and scheduled a session time. Once scheduled, they received a digital copy of the consent form.

Upon arrival, we greeted the participant at our laboratory. They signed the consent form and answered a background questionnaire on a tablet. They received general instructions, then we measured their IPD and adjusted the lenses on the Meta Quest Pro. The participant put the device on and completed the standard Meta Quest Pro eye calibration procedure. We connected the headset to the computer using a Meta Quest Link cable and started the Unity prototype. Then, we asked the participants to position the trackpad at a location that was comfortable for them and asked them to leave it there for the remainder of the study.

We calibrated the position of the virtual display, accounting for the participant's height. We then performed a short color blindness test to ensure the participant could distinguish between the four colors presented in the study. For each configuration, we provided the participants with a training session to explore the condition before the main task began. The investigator asked them to move the cursor between the four corners of the screen to get used to the size of the screen and the speed of the cursor. Then, we explained how that condition worked and let them try it out during training, where they had to complete five window switch subtasks. They moved to the main task, where they completed 55 subtasks continuously.

The participant removed the headset between each condition and answered our custom ratings questionnaire and raw NASA-TLX on an iPad. The ratings questionnaire asked participants to provide ratings for statements based on their perceptions. Once all conditions were completed, the participant answered a semi-structured interview to get some overall feedback.

\subsection{Participants}
Sixteen participants (aged 19 to 42, 6 female) from the campus population participated in the experiment in individual sessions of around 60 minutes. Twelve participants were graduate students, two were undergraduates, and two were professionals. All participants used a computer daily for work, with 12 people reporting using a computer for more than 8 hours on a typical weekday. Eleven participants rated their computer experience at 5 (out of 5), four at 4, and one at 3. All participants reported intermediate to advanced experience with the Windows operating system. The majority of participants (11) had tried VR more than 10 times, while 4 participants had tried VR at least 4 times. Eight participants rated their VR experience at 5 (out of 5), three at 4, three at 3, and two at 2. All participants reported fatigue levels lower than 3 (out of 5).

\subsection{Objective Results}
\begin{table}[tb]
\caption{Statistics of thumbnail time.}
\label{tab:ch8_1_time1}
\def\arraystretch{1.3}
\resizebox{\linewidth}{!}{%
    \begin{tabular}{llllllllll}
    \toprule
    Block   & \multicolumn{3}{l}{Selection Mode} & \multicolumn{3}{l}{Cursor Behavior} & \multicolumn{3}{l}{Interaction Effect} \\
            & $F_{1,1}$   & $p$    & $\eta_p^2$  &  $F_{1,1}$   & $p$    & $\eta_p^2$  & $F_{1,1}$   & $p$    & $\eta_p^2$ \\
    \midrule
    Overall & $12.77$       & $0.001$  & $0.004(\text{vs})$   & $10.73$   & $0.001$   & $0.003(\text{vs})$ & $28.12$   & $<0.001$   & $0.008(\text{vs})$ \\
    Large   & $90.02$       & $0.001$  & $0.05(\text{s})$     & $12.22$   & $0.001$   & $0.007(\text{vs})$ & $35.28$   & $0.001$ & $0.02(\text{s})$   \\
    Short   & $7.45$        & $0.001$  & $0.005(\text{vs})$   & --- & --- & ---   & $5.09$   & $0.024$   & $0.003(\text{vs})$  \\
    \bottomrule
    \end{tabular}
}
\vspace{-3mm}
\end{table}

\begin{figure}[tb]
 \centering
 \includegraphics[width=\columnwidth]{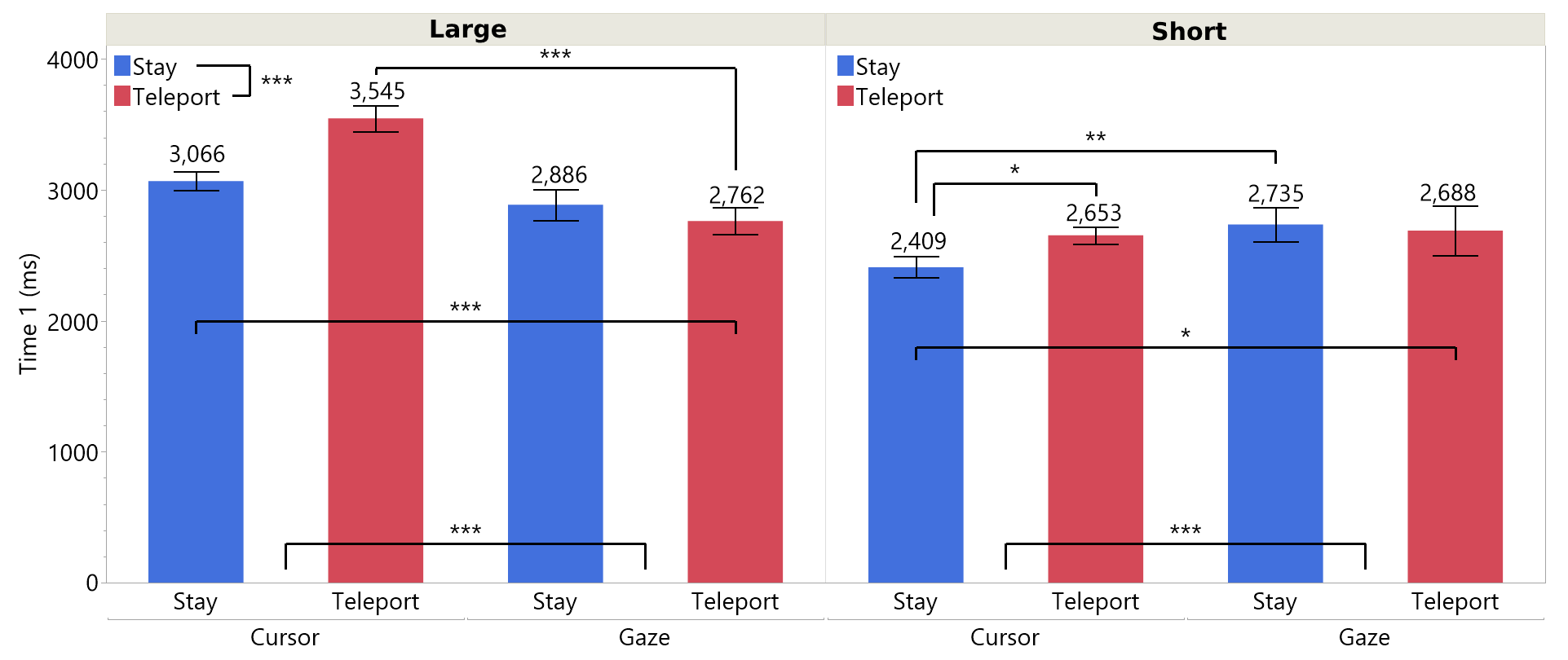}
 \caption{Thumbnail Time: large distance (left); short distance (right). Error bars represent 95\% confidence intervals.}
 \label{fig:ch8_3_time1}
 \vspace{-4mm}
\end{figure}

We collected our results from multiple sources: Unity provided time and selection errors for each window switch operation. We stored 3 times: time from the start of the task to the selection of the correct thumbnail in the spatial bar (called thumbnail time), time from the selection in the spatial bar to the selection of the next task button (called button time), and the sum of both (total time). Google Forms recorded questionnaires, including NASA TLX and statement ratings. Finally, semi-structured interviews were audio-recorded, transcribed by Office Online, and verified by the authors.

We performed the statistical analysis using the \textit{JMP Pro 16} software, with $\alpha=0.05$. Figures use * for $p \leq 0.05$, ** for $p \leq 0.01$, and *** for $p \leq 0.001$. Effect sizes are reported as partial eta squared ($\eta_p^2$) for main effects and Cohen's~$d$ for pairwise comparisons. We verified normality through Shapiro-Wilk tests and normal quantile plot inspections for all the cases before deciding whether to apply an Aligned Rank Transform (ART) \cite{wobbrock_aligned_2011} before applying a two-way analysis of variance (ANOVA). We performed pairwise comparisons using Tukey HSD (honestly significant difference) when appropriate (using values from ART-C \cite{elkin_aligned_2021} when transforming). Our two factors were the \textit{Selection Mode} (\textsc{Gaze} or \textsc{Cursor}) and \textit{Cursor Behavior} (\textsc{Teleport} or \textsc{Stay}). Aside from overall effects, we performed our analysis on blocks based on the distance to the window that impacted that measure.

\begin{table}[tb]
\caption{Statistics of button time.}
\label{tab:ch8_2_time2}
\def\arraystretch{1.3}
\resizebox{\linewidth}{!}{%
    \begin{tabular}{llllllllll}
    \toprule
    Block   & \multicolumn{3}{l}{Selection Mode} & \multicolumn{3}{l}{Cursor Behavior} & \multicolumn{3}{l}{Interaction Effect} \\
            & $F_{1,1}$   & $p$    & $\eta_p^2$  &  $F_{1,1}$   & $p$    & $\eta_p^2$  & $F_{1,1}$   & $p$    & $\eta_p^2$ \\
    \midrule
    Overall & $14.03$ & $0.001$ & $0.004(\text{vs})$ & $263.19$ & $0.001$ & $0.07(\text{m})$ & $14.28$ & $0.001$ & $0.004(\text{vs})$ \\
    Large & --- & --- & --- & $151.82$ & $0.001$ & $0.08(\text{m})$ & --- & --- & --- \\
    Short & $49.12$ & $0.001$ & $0.03(\text{s})$ & $182.94$ & $0.001$ & $0.10(\text{m})$ & $53.29$ & $0.001$ & $0.03(\text{s})$ \\
    \bottomrule
    \end{tabular}
}
\vspace{-1mm}
\end{table}

\begin{figure}[tb]
 \centering
 \includegraphics[width=\columnwidth]{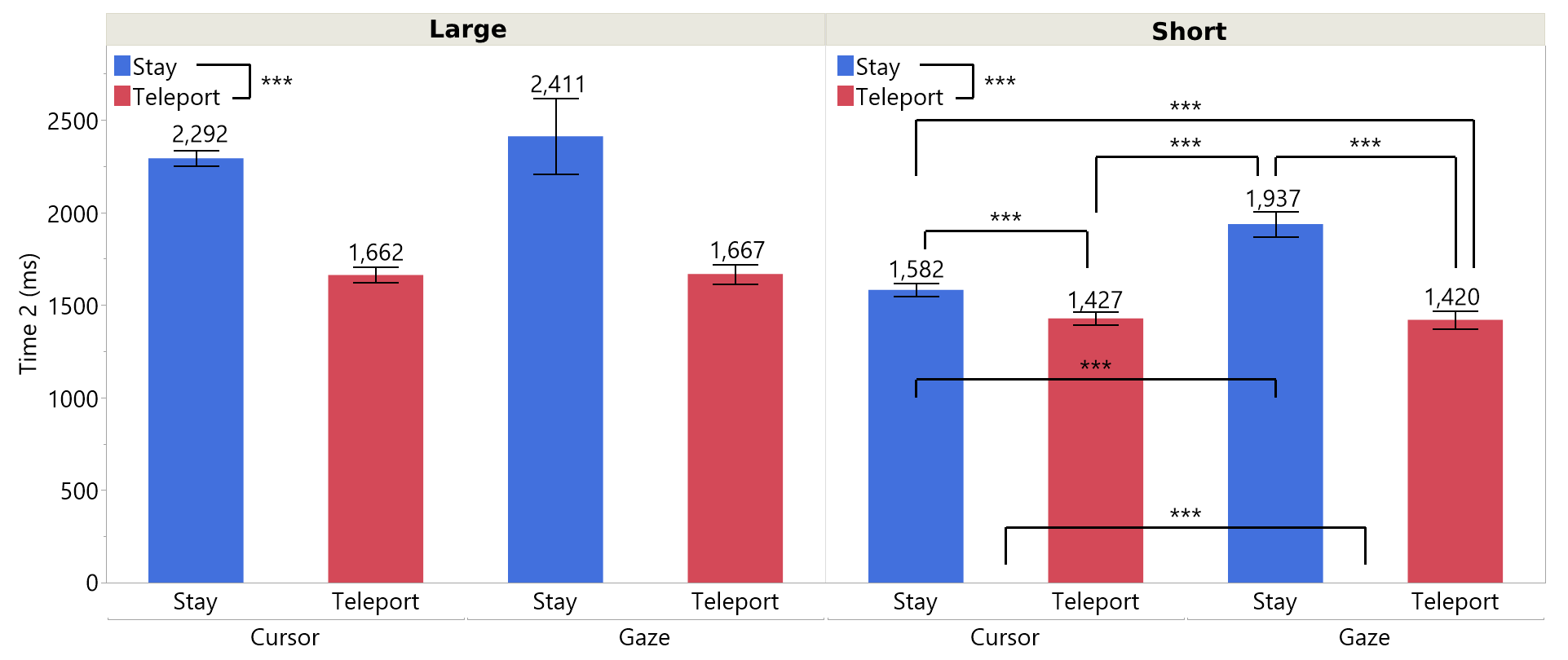}
 \caption{Button Time: large distance (left); short distance (right). Error bars represent 95\% confidence intervals.}
 \label{fig:ch8_4_time2}
 \vspace{-4mm}
\end{figure}

\subsubsection{Selection Errors}
Overall, \textsc{Cursor} yielded fewer selection errors than \textsc{Gaze} ($F_{1,1}=4.32$, $p=0.04$, $\eta_p^2=0.005$). The same was observed in the short-short block ($F_{1,1}=4.53$, $p=0.03$, $\eta_p^2=0.005$). We found no differences for the statement ``I could accurately select the desired thumbnail in the SpatialBar.''

\subsubsection{Thumbnail Time}
Overall, \textsc{Gaze} was faster than \textsc{Cursor}, and \textit{Stay} was faster than \textit{Teleport} (\autoref{tab:ch8_1_time1}). \textsc{Cursor Teleport} took longer than \textsc{Gaze Teleport} ($d=0.31$), \textsc{Cursor Stay} ($d=0.30$), and \textsc{Gaze Stay} ($d=0.24$).

\paragraph{Distance Blocks.} In \textbf{large distance}, \textsc{Gaze} was faster than \textsc{Cursor} (\autoref{fig:ch8_3_time1}). \textsc{Gaze Teleport} was faster than both \textsc{Cursor Teleport} ($d=0.76$) and \textsc{Cursor Stay} ($d=0.29$). In \textbf{short distance}, \textsc{Cursor} was faster than \textsc{Gaze}. \textsc{Cursor Stay} was faster than \textsc{Gaze Stay} ($d=0.25$), \textsc{Gaze Teleport} ($d=0.21$), and \textsc{Cursor Teleport} ($d=0.18$).

For the statement \textit{``I could quickly select the desired thumbnail in the SpatialBar,''} \textsc{Gaze} was rated higher than \textsc{Cursor} ($p=0.074$).

\subsubsection{Button Time}
Overall, \textsc{Cursor} was faster than \textsc{Gaze}, and \textsc{Teleport} was faster than \textsc{Stay} (\autoref{tab:ch8_2_time2}). \textsc{Gaze Teleport} was faster than \textsc{Gaze Stay} ($d=0.69$) and \textsc{Cursor Stay} ($d=0.43$). \textsc{Cursor Teleport} also outperformed both \textsc{Gaze Stay} ($d=0.69$) and \textsc{Cursor Stay} ($d=0.43$). \textsc{Cursor Stay} was faster than \textsc{Gaze Stay} ($d=0.26$).

\paragraph{Distance Blocks.} In \textbf{large distance}, \textsc{Teleport} was faster than \textsc{Stay} (\autoref{fig:ch8_4_time2}). In \textbf{short distance}, \textsc{Cursor} was faster than \textsc{Gaze}, and \textsc{Teleport} was faster than \textsc{Stay}. \textsc{Gaze Teleport} was faster than \textsc{Gaze Stay} ($d=1.02$) and \textsc{Cursor Stay} ($d=0.32$). Likewise, \textsc{Cursor Teleport} outperformed \textsc{Gaze Stay} ($d=1.02$) and \textsc{Cursor Stay} ($d=0.32$). Finally, \textsc{Cursor Stay} was faster than \textsc{Gaze Stay} ($d=0.70$).

In \textit{``I could quickly press the button in the window after selecting the thumbnail,''} \textsc{Teleport} was rated significantly higher than \textsc{Stay} ($F_{1,1}=11.94$, $p=0.001$, $\eta_p^2=0.16$).

\subsubsection{Total Time}
\begin{table}[tb]
\caption{Statistics of total time.}
\label{tab:ch8_3_totalTime}
\def\arraystretch{1.3}
\resizebox{\linewidth}{!}{%
    \begin{tabular}{llllllllll}
    \toprule
    Block   & \multicolumn{3}{l}{Selection Mode} & \multicolumn{3}{l}{Cursor Behavior} & \multicolumn{3}{l}{Interaction Effect} \\
            & $F_{1,1}$   & $p$    & $\eta_p^2$  &  $F_{1,1}$   & $p$    & $\eta_p^2$  & $F_{1,1}$   & $p$    & $\eta_p^2$ \\
    \midrule
    Overall & --- & --- & --- & $44.55$ & $0.001$ & $0.023(\text{s})$ & $37.48$ & $0.001$ & $0.011(\text{s})$ \\
    LL & $9.52$ & $0.002$ & $0.01(\text{s})$ & $21.22$ & $0.001$ & $0.023(\text{s})$ & $10.30$ & $0.001$ & $0.011(\text{s})$ \\
    LS & $6.86$ & $0.009$ & $0.008(\text{vs})$ & $10.45$ & $0.001$ & $0.012(\text{s})$ & $53.42$ & $0.001$ & $0.061(\text{m})$ \\
    SL & $7.34$ & $0.007$ & $0.008(\text{vs})$ & $24.80$ & $0.001$ & $0.027(\text{s})$ & --- & --- & --- \\
    SS & --- & --- & --- & --- & --- & ---  & $3.96$ & $0.047$ & $0.004(\text{vs})$ \\
    \bottomrule
    \end{tabular}
}
\vspace{-3mm}
\end{table}

Overall, \textsc{Teleport} was faster than \textsc{Stay} \autoref{tab:ch8_3_totalTime}. Pairwise comparisons showed \textsc{Gaze Teleport} outperformed \textsc{Gaze Stay} ($d=0.44$), \textsc{Cursor Stay} ($d=0.25$), and \textsc{Cursor Teleport} ($d=0.23$).

\begin{figure}[tb]
 \centering
 \includegraphics[width=\columnwidth]{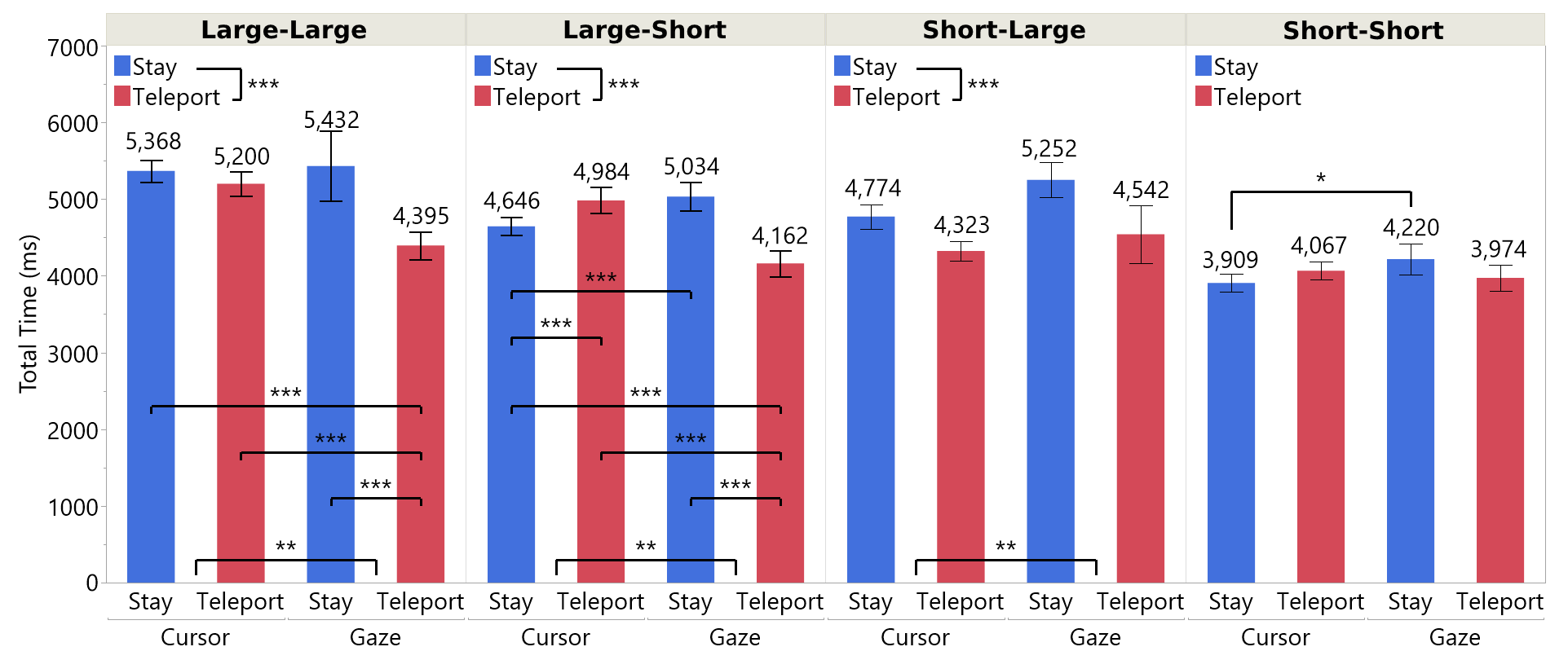}
 \caption{Total Time in distance blocks: Large-Large (left); Large-Short; Short-Large; Short-Short (right). Error bars represent 95\% confidence intervals.}
 \label{fig:ch8_4_totalTime}
 \vspace{-5mm}
\end{figure}

\paragraph{Distance Blocks.} In the \textbf{large-large block}, \textsc{Gaze} was faster than \textsc{Cursor}, and \textsc{Teleport} was faster than \textsc{Stay} (\autoref{fig:ch8_4_totalTime}). \textsc{Gaze Teleport} was faster than \textsc{Gaze Stay}  ($d=0.53$), \textsc{Cursor Stay} ($d=0.49$), and \textsc{Cursor Teleport} ($d=0.41$).
In the \textbf{large-short block} \textsc{Gaze} was faster than \textsc{Cursor}, and \textsc{Teleport} was faster than \textsc{Stay}. \textsc{Gaze Teleport} significantly outperformed the other LS conditions ($p<0.001$), and \textsc{Cursor Stay} led to less time than both \textsc{Gaze Stay} and \textsc{Cursor Teleport}.
In the \textbf{short-large block}, \textsc{Cursor} was faster than \textsc{Gaze}) and \textsc{Teleport} was faster than \textsc{Stay}.
In the \textbf{short-short block}, \textsc{Cursor Stay} was faster than \textsc{Gaze Stay} ($d=0.24$).

In ``I could switch between windows quickly'', participants favored \textsc{Gaze} over \textsc{Cursor} ($F_{1,1}=6.33$, $p=0.014$, $\eta_p^2=0.09$), and \textsc{Teleport} over \textsc{Stay} ($F_{1,1}=6.13$, $p=0.016$, $\eta_p^2=0.09$).

\subsubsection{Task Load}
Regarding \textbf{physical demand}, \textsc{Teleport} was rated lower in physical demand than \textsc{Stay} ($F_{1,1}=6.52$, $p=0.013$, $\eta_p^2=0.095$). No additional significant effects were found.

For the statement \textit{``I did not have to move the cursor a lot,''} \textsc{Gaze} scored higher than \textsc{Cursor} ($F_{1,1}=28.58$, $p<0.001$, $\eta_p^2=0.32$), and \textsc{Teleport} outperformed \textsc{Stay} ($F_{1,1}=37.96$, $p<0.001$, $\eta_p^2=0.38$). An interaction effect ($F_{1,1}=4.01$, $p=0.049$, $\eta_p^2=0.06$) further highlighted \textsc{Gaze Teleport} as significantly higher than \textsc{Cursor Stay} ($d=2.66$), \textsc{Gaze Stay} ($d=2.15$), and \textsc{Cursor Teleport} ($d=1.79$).

\subsection{Qualitative Results}

\subsubsection{\textsc{Cursor Stay} was familiar and comfortable, but laborious and took longer}
Seven participants found the \textsc{Cursor Stay}  technique to be the most familiar and comfortable. P0 remarked, ``cursor stay was the most familiar one.'' P4 stated, ``Gaze teleport and cursor stay were easier for me... I just found the cursor was easier to move and I didn't struggle locating the cursor.'' P7 added, ``cursor stay was what I was most comfortable with... I think I moved the fastest because I was used to it.'' P8 further mentioned, ``I just need to move it directly to where it is instead of having to think if it is going to teleport and try to compensate for that.''

While participants found \textsc{Cursor Stay}  familiar, five described it as laborious and time-consuming. P0 said, ``cursor stay was really slow, or it felt slower because of how much movement I am doing.'' P1 highlighted the tedium involved, stating, ``in cursor stay, having to move the cursor from a window all the way down to the taskbar, then all the way up to a window... repeating that got tedious and felt less intuitive.'' P13 summarized the general sentiment by saying, ``we're used to using cursor stay, but it felt very laborious to use that interface.''

\subsubsection{\textsc{Gaze Teleport} was easy, intuitive, and perceived as fastest, although with a learning curve}
Thirteen participants praised \textsc{Gaze Teleport} as the most intuitive and efficient interaction technique. P7 appreciated its simplicity, stating, ``definitely gaze teleport, as it was really, really easy to just switch windows, especially with the large screen that there was.'' P10 emphasized its ease of use and productivity benefits: ``gaze teleport was less demanding and I could just look at what I wanted, and I just needed to click, and it teleport to the center of the window.'' Eight participants perceived \textsc{Gaze Teleport} as the fastest condition. P0 stated, ``of course gaze teleport was definitely the fastest.'' P6 emphasized, ``The gaze was much faster because I didn't have to then move the cursor to the thumbnail to the space bar.''

Four participants mentioned that \textsc{Gaze Teleport} had a learning curve. P2 said, ``As time went on, despite the tracking not being perfect, I started to like gaze more.'' P10 agreed, ``At first, it was challenging to get used to how gaze teleport worked as I didn’t need to move the cursor a lot, but moving forward, it was the easiest one.'' P14 summarized, ``Gaze Teleport was convenient… it had a hard learning curve, so it might have taken more time to get used to it, but when you do, it was more convenient because it was similar to when you use two hands to do something.''

\subsubsection{\textsc{Cursor Teleport} was a compromise but inconsistent}
Five participants found \textsc{Cursor Teleport} to be a compromise. P3 described it neutrally, ``cursor teleport was also not bad.'' P10 noted, ``It was not the traditional way of working with the cursor, but it made it a bit easier. It was more interesting than the cursor stay.'' P12 said, ``cursor teleport felt good because it was a good blending to me. I found it to be less rushed and to take less mental demand because the other ones I was a little bit more like rushing and I tried to cope with the pace of the movement.''

However, six other participants found \textsc{Cursor Teleport} inconsistent. P0 described, ``Cursor teleport was really confusing because… the window opens here, and I'm still in the mindset of needing to move the cursor there… but the cursor already teleported there.'' P15 complemented, ``I would prefer cursor stay to cursor teleport because I ended up wasting that teleport every time anyway.'' P0 suggested a preference for more automated systems: ``I think you either automate everything or you don't. The inconsistency was really annoying… so either gaze teleport or cursor stay.''

\subsubsection{\textsc{Gaze Stay} required modality switching and memory}
Four participants highlighted challenges with \textsc{Gaze Stay} in switching between the modalities and remembering the cursor location. P10 mentioned, ``I would get confused a few times because the movement of the hand and the movement of the gaze… there were two different mechanisms, and I had to try to get them to contribute together.'' P6 said, ``It wasn't as confusing in the gaze teleport, because once I selected, I knew that I just needed to move the cursor from the middle of the window to where the button is. But then with the gaze stay, I might have to start looking for the cursor.'' 

\subsubsection{\textsc{Cursor} feels more precise and deliberate}
Three participants found using the \textsc{Cursor} to be more precise and deliberate compared to gaze. P1 said, ``I still think my brain had to process where to look, whereas moving my cursor don't take a lot of time because of how close they are.'' P11 stated, ``with the cursor it was much more deliberate, I was much more accurate choosing which thumbnail I wanted... with the cursors if I forgot the next target, I could just look back up at the number at the same time I was moving the cursor down.'' P13 said, ``I feel like using the cursor ones I was able to select things better.''

\subsubsection{\textsc{Gaze} required users to deal with two modalities}
Four participants mentioned the challenge of dealing with two modalities. P6 noted, "It was challenging to switch between using my eyes to select the thumbnails and then moving the cursor to button in the window. From just the choice of using my eyes or using the cursor back and forth." P13 highlighted a challenge, "as soon as there was a distraction, I sort of forgot how to interact with it ... I think the challenge I faced was to remember what interface I'm using right now." P14 mentioned, "It took some time to getting used to the gaze, because initially I'd move my cursor with it, but because we were using both gaze and cursor separately."

\subsubsection{Eye tracking can be imprecise and fatiguing}
Four participants mentioned the eye tracking being imprecise. P7 said, ``I would say the biggest nuisance is the eye tracker in the headset itself. There are times where I felt that I was looking at one of the thumbnails but it didn't lock on to the thumbnail.'' P11 complained about thumbnails getting highlighted when looked at, ``the highlight made me look at it sometimes because it flashes white and you want to look at it… it's grabbing my attention, so I look at it and then sometimes I misclicked.'' Two participants further mentioned their eyes getting more strained during \textsc{Gaze} conditions.

\section{Discussion}
Our first hypothesis stated that \textsc{Gaze} would perform faster than \textsc{Cursor} when the previous window location was farther from the spatial bar (\textbf{H1}). Our results support this hypothesis. We found a statistically significant difference with \textsc{Gaze} leading to 14\% shorter time than \textsc{Cursor}. Further, 81\% of participants mentioned that \textsc{Gaze Teleport} was the easiest or most intuitive condition they tried. Another eight participants mentioned that \textsc{Gaze Teleport} was perceived as the fastest.

Our second hypothesis stated that \textsc{Cursor} would be faster than \textsc{Gaze} when the previous window location was closer to the spatial bar (\textbf{H2}). We found evidence to support this hypothesis. For short distances, we found that using \textsc{Cursor} led to 10\% shorter times than using \textsc{Gaze}, which was significant. This helps us understand the overhead of switching between using two modalities to control selection. When the distance is large, the benefit of using \textsc{Gaze} is larger than the cost, which is not true for the short distances we used in this study.

Our third hypothesis stated that \textsc{Gaze} would lead to a higher selection error rate than \textsc{Cursor}. We can only partially support this hypothesis (\textbf{H3}). While \textsc{Cursor} overall had fewer errors than \textsc{Gaze}, we only obtained statistical significance in the Short-Short block, when the task started from a window close to the spatial bar and ended at another window that was also close by. In this block, \textsc{Gaze} had 4x more errors than \textsc{Cursor}. However, the overall error rate was still small for both (2.2\% and 0.4\%, respectively). From the fact that those results were not repeated for other distance blocks, we could speculate two possibilities: that users were actively trying to make the window switch faster at shorter distances, and thus incurring more errors; or it could be related to imprecision of eye tracking, since for short-distance trials, participants may have tended to move only their eyes, not their heads, and eye tracking may be less precise in the periphery.

The fourth hypothesis stated that \textsc{Gaze Teleport} would lead to a decrease in task completion time and workload compared to \textsc{Gaze Stay} (\textbf{H4}). Our results provide evidence to partially support this hypothesis. The pairing of \textsc{Gaze Teleport} yielded the best speed performance for large distances, while \textsc{Gaze Stay} was the worst performing technique at all distances. As \textsc{Gaze} introduces a cost in switching modalities, its benefits in performance must outweigh such costs. \textsc{Gaze Stay} didn't reach such a threshold, not only penalizing users for the modality switch but also requiring them to keep track of the previous cursor location or locate it in case it went missing. Regarding NASA TLX, we didn't find any evidence to support any difference in task load between the conditions.

Our fifth hypothesis stated that \textsc{Cursor Stay} would perform better than \textsc{Cursor Teleport}, leading to lower completion times and cognitive task load  (\textbf{H5}). We cannot support this hypothesis. There is evidence that \textsc{Teleport} outperforms \textsc{Stay} in most situations. When we look at pairwise comparisons, however, most data indicates no difference between \textsc{Cursor Stay} and \textsc{Cursor Teleport}. Regarding NASA TLX, we found evidence that \textsc{Teleport} led to a lower physical demand than \textsc{Stay}, but those were not found in other metrics and were not specific to pairing with \textsc{Cursor}.

Overall, we can provide recommendations: \textsc{Gaze Teleport} can be faster and should be preferred for large displays where windows have large distances, while \textsc{Cursor Stay} should be used when targets are close together. \textsc{Gaze} can be less accurate due to imprecision, and \textsc{cursor} feels more precise and deliberate---that cost should be considered when adopting it. If using \textsc{gaze} or \textsc{teleport}, they should be paired together, as \textsc{Gaze Stay} can require users to remember cursor position and make it harder to switch modality while \textsc{Cursor Teleport} can be perceived as inconsistent, where the user must assimilate behavior and not overshoot.

\section{Limitations}
This work includes some limitations. Our study included a reduced number of participants, and such findings should be replicated. We also focused on the specific window-switching task and did not include more realistic windows and window placements. Our results could further be dependent on the level of eye tracking technology, which might improve the future, and the size of the buttons selected using the gaze. Finally, results might be different using a physical monitor and mouse instead of a virtual monitor and trackpad, and a follow-up investigation on those conditions should be done.

\section{Conclusions and Future Work}
In this paper, we contribute the Spatial Bar interface for window switching on large virtual displays. It shows all open windows in the system as thumbnails presented in an underutilized space below the virtual display. We conducted a user study to understand the effect of design choices on performance and cognitive load. Results indicated that \textsc{Gaze} was faster than \textsc{Cursor} when the target window was farther from the spatial bar. Conversely, \textsc{Cursor} outperformed \textsc{Gaze} when the target window was closer. \textsc{Gaze} exhibited a higher error rate in scenarios where windows were close to the Spatial Bar. \textsc{Teleport} in general reduced task completion times by minimizing cursor travel. \textsc{Gaze Teleport} worked well for large distances, but \textsc{Cursor Teleport} required users to adapt to the teleportation mechanism to avoid overshooting or confusion in cursor location. Finally, in \textsc{Cursor Stay}, users experienced familiarity and comfort, but at the cost of a more laborious and slower interaction.

Our novel \textsc{Gaze Teleport} design proved effective for managing window switches in large displays, where the user would need to move large distances to select their active window. It does not seem effective for more traditional monitors, where distances are shorter and less time-consuming. It shows that the combination of gaze and cursor modalities can be effective, but we must be careful about the base cost of switching between the two modalities. One important aspect to be highlighted after this study is the importance of cues to help users transition between the modalities. \textsc{Gaze Stay}  performed worse than \textsc{Gaze Teleport} because it required users to perform a broader context switch, having to remember where the cursor was on top of remembering to switch from using gaze to cursor. In this study, we used the animation cue for all conditions, as we already considered it essential to support such a switch from the beginning. In the future, it would be interesting to investigate what other cues could be used to support such transitions and a more broad and systematic definition of when each modality could be used.

\bibliographystyle{abbrv-doi}

\bibliography{references}
\end{document}


\makeatletter
\setlength{\@fptop}{0pt}
\makeatother
\begin{table}[H]
\centering
\caption{Descriptive statistics for thumbnail, button, and total time. All reported means ($\mu$) and standard deviations ($\sigma$).}
\label{tab:allResults}
\def\arraystretch{1.2}
\resizebox{!}{.4\paperheight}{%
\begin{tabular}{llll}
\toprule
\textbf{Block} & \textbf{Condition} & $\mu$ & $\sigma$ \\
\midrule
\multicolumn{4}{c}{\textbf{Thumbnail Time}} \\
\midrule
\multirow{8}{*}{\textbf{Overall}} & \textsc{Gaze} & 2767 & 1450 \\
 & \textsc{Cursor} & 2918 & 939 \\
 & \textsc{Stay} & 2773 & 1106 \\
 & \textsc{Teleport} & 2911 & 1328 \\
 & \textsc{Cursor Teleport} & 3098 & 978 \\
 & \textsc{Gaze Teleport} & 2724 & 1582 \\
 & \textsc{Cursor Stay} & 2737 & 861 \\
 & \textsc{Gaze Stay} & 2810 & 1305 \\
\midrule
\multirow{7}{*}{\textbf{Large distance}} & \textsc{Gaze} & 2823 & 1154 \\
 & \textsc{Cursor} & 3305 & 932 \\
 & \textsc{Stay} & 2975 & 1020 \\
 & \textsc{Teleport} & 3153 & 1123 \\
 & \textsc{Gaze Teleport} & 2761 & 1077 \\
 & \textsc{Cursor Teleport} & 3544 & 1029 \\
 & \textsc{Cursor Stay} & 3065 & 753 \\
\midrule
\multirow{6}{*}{\textbf{Short distance}} & \textsc{Cursor} & 2530 & 772 \\
 & \textsc{Gaze} & 2711 & 1694 \\
 & \textsc{Cursor Stay} & 2408 & 838 \\
 & \textsc{Gaze Stay} & 2734 & 1378 \\
 & \textsc{Gaze Teleport} & 2687 & 1961 \\
 & \textsc{Cursor Teleport} & 2652 & 678 \\
\midrule
\multicolumn{4}{c}{\textbf{Button Time}} \\
\midrule
\multirow{8}{*}{\textbf{Overall}} & \textsc{Cursor} & 1740 & 519 \\
 & \textsc{Gaze} & 1858 & 1233 \\
 & \textsc{Teleport} & 1544 & 480 \\
 & \textsc{Stay} & 2055 & 1198 \\
 & \textsc{Gaze Teleport} & 1543 & 537 \\
 & \textsc{Gaze Stay} & 2174 & 1599 \\
 & \textsc{Cursor Stay} & 1936 & 538 \\
 & \textsc{Cursor Teleport} & 1544 & 415 \\
\midrule
\multirow{2}{*}{\textbf{Large distance}} & \textsc{Teleport} & 1652 & 489 \\
 & \textsc{Stay} & 2330 & 1532 \\
\midrule
\multirow{8}{*}{\textbf{Short distance}} & \textsc{Cursor} & 1504 & 373 \\
 & \textsc{Gaze} & 1678 & 667 \\
 & \textsc{Teleport} & 1423 & 439 \\
 & \textsc{Stay} & 1759 & 594 \\
 & \textsc{Gaze Teleport} & 1420 & 504 \\
 & \textsc{Gaze Stay} & 1937 & 709 \\
 & \textsc{Cursor Stay} & 1581 & 369 \\
 & \textsc{Cursor Teleport} & 1427 & 362 \\
\midrule
\multicolumn{4}{c}{\textbf{Total Time}} \\
\midrule
\multirow{6}{*}{\textbf{Overall}} & \textsc{Teleport} & 4455 & 1507 \\
 & \textsc{Stay} & 4829 & 1731 \\
 & \textsc{Gaze Teleport} & 4268 & 1774 \\
 & \textsc{Gaze Stay} & 4984 & 2170 \\
 & \textsc{Cursor Stay} & 4674 & 1112 \\
 & \textsc{Cursor Teleport} & 4643 & 1152 \\
\midrule
\multirow{8}{*}{\textbf{Large-Large}} & \textsc{Gaze}            & 4913 & 2603 \\
 & \textsc{Cursor}          & 5283 & 1091 \\
 & \textsc{Teleport}        & 4797 & 1293 \\
 & \textsc{Stay}            & 5399 & 2486 \\
 & \textsc{Gaze Teleport}   & 4395 & 1317 \\
 & \textsc{Gaze Stay}       & 5431 & 3363 \\
 & \textsc{Cursor Teleport} & 5199 & 1137 \\
 & \textsc{Cursor Stay}     & 5368 & 1039 \\
\midrule
\multirow{8}{*}{\textbf{Large-Short}} & \textsc{Gaze}            & 4598 & 1366 \\
 & \textsc{Cursor}          & 4814 & 1092 \\
 & \textsc{Teleport}        & 4572 & 1317 \\
 & \textsc{Stay}            & 4840 & 1145 \\
 & \textsc{Gaze Teleport}   & 4162 & 1250 \\
 & \textsc{Gaze Stay}       & 5034 & 1340 \\
 & \textsc{Cursor Teleport} & 4983 & 1256 \\
 & \textsc{Cursor Stay}     & 4646 & 871  \\
\midrule
\multirow{4}{*}{\textbf{Short-Large}} & \textsc{Cursor}          & 4548 & 1064 \\
 & \textsc{Gaze}            & 4897 & 2310 \\
 & \textsc{Teleport}        & 4432 & 2061 \\
 & \textsc{Stay}            & 5013 & 1453 \\
\midrule
\multirow{2}{*}{\textbf{Short-Short}} & \textsc{Cursor Stay} & 3908 & 834 \\
 & \textsc{Gaze Stay} & 4219 & 1474 \\
\bottomrule
\end{tabular}
}
\end{table}

\begin{table}[H]
\centering
\caption{Descriptive statistics for selection errors. All reported means ($\mu$) and standard deviations ($\sigma$).}
\label{tab:allResults}
\def\arraystretch{1.2}
\begin{tabular}{llll}
\toprule
\textbf{Block} & \textbf{Condition} & $\mu$ & $\sigma$ \\
\midrule
\multirow{2}{*}{\textbf{Overall}} & \textsc{Cursor} & 0.009 & 0.09 \\
 & \textsc{Gaze} & 0.017 & 0.13 \\
\multirow{2}{*}{\textbf{Short-Short}} & \textsc{Cursor} & 0.005 & 0.07 \\
 & \textsc{Gaze} & 0.02 & 0.14 \\
\bottomrule
\end{tabular}
\end{table}

\begin{table}[H]
\centering
\caption{Descriptive statistics for physical demand. All reported means ($\mu$) and standard deviations ($\sigma$).}
\label{tab:allResults}
\def\arraystretch{1.2}
\begin{tabular}{llll}
\toprule
\textbf{Block} & \textbf{Condition} & $\mu$ & $\sigma$ \\
\midrule
\multirow{2}{*}{\textbf{Overal}} & \textsc{Teleport} & 2.22 & 1.41 \\
 & \textsc{Stay} & 3.31 & 1.80 \\
\end{tabular}
\end{table}